\pdfoutput=1

\documentclass[10pt,journal]{IEEEtran}
\usepackage{amsmath,amsfonts}
\usepackage{algorithmic}
\usepackage{algorithm}
\usepackage{array}
\usepackage[caption=false,font=normalsize,labelfont=sf,textfont=sf]{subfig}
\usepackage{textcomp}
\usepackage{stfloats}
\usepackage{url}
\usepackage{verbatim}
\usepackage{graphicx}
\usepackage{cite}
\usepackage{xcolor}

\usepackage{multirow}
\begin{document}
\title{EMOVOME: A Dataset for Emotion Recognition in Spontaneous Real-Life Speech}

\author{Lucía~Gómez-Zaragozá, Rocío~del~Amor, María~José~Castro-Bleda, Valery~Naranjo, Mariano~Alcañiz~Raya, Javier~Marín-Morales
        % <-this % stops a space
% \thanks{This work has been submitted to the IEEE for possible publication. Copyright may be transferred without notice, after which this version may no longer be accessible.} % for arXiv
\thanks{This paper was produced by the IEEE Publication Technology Group. They are in Piscataway, NJ.}% <-this % stops a space
\thanks{Manuscript received November 29, 2024; revised XXXX XX, 2024.}
}

% \markboth{ }% % for arXiv
\markboth{Journal of \LaTeX\ Class Files,~Vol.~14, No.~8, August~2021}%
{Shell \MakeLowercase{\textit{et al.}}: A Sample Article Using IEEEtran.cls for IEEE Journals}


\maketitle

% ##########################################################################################
% Abstract
% ##########################################################################################
\begin{abstract} 
Spontaneous datasets for Speech Emotion Recognition (SER) are scarce and frequently derived from laboratory environments or staged scenarios, such as TV shows, limiting their application in real-world contexts. We developed and publicly released the Emotional Voice Messages (EMOVOME) dataset, including 999 voice messages from real conversations of 100 Spanish speakers on a messaging app, labeled in continuous and discrete emotions by expert and non-expert annotators. We evaluated speaker-independent SER models using acoustic features as baseline and transformer-based models. We compared the results with reference datasets including acted and elicited speech, and analyzed the influence of annotators and gender fairness. The pre-trained UniSpeech-SAT-Large model achieved the highest results, 61.64\% and 55.57\% Unweighted Accuracy (UA) for 3-class valence and arousal prediction respectively on EMOVOME, a 10\% improvement over baseline models. For the emotion categories, 42.58\% UA was obtained. EMOVOME performed lower than the acted RAVDESS dataset. The elicited IEMOCAP dataset also outperformed EMOVOME in predicting emotion categories, while similar results were obtained in valence and arousal. EMOVOME outcomes varied with annotator labels, showing better results and fairness when combining expert and non-expert annotations. This study highlights the gap between controlled and real-life scenarios, supporting further advancements in recognizing genuine emotions.

\end{abstract}

% ##########################################################################################
% Keywords
% ##########################################################################################

% ##########################################################################################
% Introduction
% ##########################################################################################

\section{Introduction}
\label{sec:introduction}

\IEEEPARstart{H}{uman} communication conveys not only ideas but also emotional states, which play a crucial role in shaping interpersonal interactions. In everyday conversations, individuals rely on both verbal content and emotional cues to interpret meaning and adjust their responses, highlighting the significance of emotion recognition. Speech Emotion Recognition (SER) is an evolving research field aimed at automatically identifying a speaker's emotional state from their voice. With applications ranging from detecting emotional distress in customer service interactions to enhancing naturalness in speech synthesis systems, SER holds significant promise for improving both human-to-human and human-computer interactions. Central to SER are emotional datasets, which are categorized in the literature based on various criteria, including the emotion models used for labeling, the authenticity of emotional expressions, and the recording environment.

The labeling of emotions in emotional datasets is influenced by the underlying emotion model used, which can be either discrete or continuous. Grounded in basic emotion theories, discrete models consider emotions as universal and distinct categories, including happiness, sadness, and anger among others \cite{ekman1999}. In contrast, dimensional models, based on core affect theory, represent emotions as points within a continuous space defined by affective dimensions such as valence and arousal \cite{russell1980circumplex}. The choice of an emotion model directly impacts the scope and granularity of the dataset, shaping its suitability for different SER applications.

Emotional datasets are also distinguished by the expression authenticity of their samples. In the literature, they are commonly classified as acted, elicited, and spontaneous \cite{swain2018, akccay2020, madanian2023}. Acted (or simulated) datasets comprise speech samples of actors simulating emotions, typically using predetermined texts. Acted emotions tend to be stereotypical and exaggerated, limiting their applicability in real-life scenarios. Elicited (or induced) datasets gather speech samples from artificial situations designed to elicit specific emotional states, such as listening to a story or watching a video. The induction process has limitations and ethical implications, as individuals may react differently to the same stimulus. Spontaneous (or natural) datasets aim to capture genuine emotional expression that occurs without external guidance or deliberate performance, making them the most challenging to collect.

The recording environment refers to the setting in which emotional data is collected, encompassing the level of control and external influences during the recording process. It can be categorized into staged, laboratory, and real-life settings. Staged settings involve controlled scenarios with carefully designed setups, such as studios for TV shows or YouTube videos, which pose certain limitations. The artificial situation may inadvertently influence speakers, leading to controlled or unnatural emotional expressions \cite{mcintyre2008composite}. Additionally, the monologue format of some YouTube videos and podcasts may lack the naturalness typically found in conversational interactions \cite{ELAYADI2011572}. It is also worth noting that the individuals present in these contexts are generally communication experts, limiting the extrapolation to the general population. Laboratory settings provide controlled environments with some degree of spontaneity when used to simulate real-life scenarios, although they can also influence the authenticity of expressions \cite{mcintyre2008composite}. Exceptions to these limitations exist in datasets from real-life scenarios, such as call center records. These datasets prioritize ecological validity but are typically restricted and come with challenges such as background noise and variability in recording conditions, called in-the-wild conditions. 

SER is still an open-ended problem due to its complexity, particularly when addressing spontaneous emotions in real-life environments. Collecting spontaneous emotional data is inherently difficult and relies on subjective external evaluation to identify the emotion in each sample, a process complicated by the variability in emotional perception among raters \cite{ding2022}. Furthermore, real-life settings are scarce in literature and often restricted from public use due to ethical and legal considerations. As a result, while most emotional corpora are predominantly in English, owing to its global prevalence \cite{akccay2020}, the availability of natural datasets remains limited. Three exceptions including spontaneous emotions from real-life scenarios were found in the literature: NATURAL \cite{MORRISON200798}, Lee \& Narayanan \cite{min2005} and SUSAS \cite{hansen1997getting} datasets. The first two include call-center recordings but have restricted public access. SUSAS includes a commercially available subset, but it is limited to recordings of isolated words from aircraft communications. The availability of emotional datasets is considerably limited for languages other than English, including Spanish. We found only two spontaneous datasets in Spanish: MOUD \cite{rosas2013multimodal} and CMU-MOSEAS \cite{zadeh2020}. Although both datasets are publicly available, they consist of monologue clips sourced from YouTube, thus recorded in staged controlled settings. The limited availability of natural datasets from real-life scenarios restricts the advancement of SER models in real-world settings.

To fill the gap in existing literature on emotional datasets with spontaneous emotions from real-life scenarios, we collected the Emotional Voice Messages (EMOVOME) dataset. It contains 999 audio messages collected from real WhatsApp conversations of 100 Spanish speakers. To the best of our knowledge, this is the first public dataset with spontaneous emotions from conversations collected in a real-life scenario. We created SER models using acoustic features and speech embeddings derived from state-of-the-art pre-trained models. To simulate more realistic and challenging conditions, we assessed the models under a speaker-independent approach, using different speakers for training and evaluation. We conducted a comprehensive analysis to assess how various EMOVOME properties impact on SER model results. We analyzed the influence of the expert and non-expert annotators, and evaluated model fairness, specifically assessing differences in performance outcomes based on the speakers' gender. We compared the results with the widely used IEMOCAP \cite{busso2008iemocap} and RAVDESS \cite{livingstone2018} datasets. 

The main contributions of the article are: (1) presenting EMOVOME, the first public dataset with spontaneous real-life emotions for SER, and making it available to the scientific community; (2) analyzing the SER performance of the state-of-the-art pre-trained speech models in real-life spontaneous conversations, and comparing the results with two non-natural reference datasets; (3) analyzing the influence of EMOVOME annotators on SER results; and (4) evaluating models' gender fairness. This study emphasizes the disparity between staged or laboratory settings and real-life scenarios, aiding in the advancement of recognizing authentic emotions.

The paper is organized as follows. Section 2 reviews existing SER datasets and related works. Section 3 describes EMOVOME and Section 4 other datasets used. Section 5 explains the methodology for creating SER models, and Section 6 reports the results. Section 7 discusses the results, and the paper concludes with a summary and future directions.

% ##########################################################################################
% Related work
% ##########################################################################################
\section{Related work}

% -------------------------------------------------------------------------
%  Existing SER datasets
% -------------------------------------------------------------------------
\subsection{Spontaneous datasets for SER}

The availability of spontaneous datasets in the literature is limited. We conducted a search using KAPODI \cite{diconne2022presenting}, a recently published searchable database of emotional stimulus sets, along with a review of the relevant literature. Table \ref{tab:databases_english} presents the spontaneous datasets identified for English and Spanish, indicating the recording environment (RE) and its details, and whether they are free to access.

\begin{table}[!ht]
    \caption{Spontaneous datasets for SER in English and Spanish. The asterisk (*) indicates if the dataset is multilingual.}
    \label{tab:databases_english}
    \centering
    \begin{tabular}{lllll} \hline
        Dataset & RE & Details & Free  \\ \hline
        \textit{English} \\ \hline
        AFEW \cite{dhall2011acted} & Staged &  TV shows & Yes  \\ 
        Belfast Naturalistic \cite{douglas2000new} & Staged & TV shows & Yes \\ 
        Castaway Reality TV \cite{douglas2007humaine} & Staged & TV shows & Yes \\ 
        MSP-PODCAST \cite{lotfian2019} & Staged & Podcasts & Yes \\ 
        AVIC \cite{schuller2009being} & Lab & Guided interaction & Yes \\ 
        NATURAL \cite{MORRISON200798} & Real-life & Call-center (human) & No \\ 
        Lee \& Narayanan \cite{min2005} & Real-life & Call-center (machine)  & No  \\ 
        SUSAS \cite{hansen1997getting} & Real-life & Pilot communications & Yes \\ 
         & & Doctor-patient interview & No \\  \hline
        \textit{Spanish} \\ \hline
        MOUD \cite{rosas2013multimodal} & Staged &  YouTube videos & Yes  \\ 
        CMU-MOSEAS * \cite{zadeh2020} & Staged & YouTube videos & Yes \\ \hline
    \end{tabular}
\end{table}

For English, we identified four \cite{dhall2011acted, douglas2000new, douglas2007humaine, lotfian2019} spontaneous datasets recorded in staged scenarios, primarily featuring clips from TV shows, and one \cite{schuller2009being} recorded in a laboratory setting, capturing interactions during a simulated product presentation. Regarding real-life scenarios, NATURAL \cite{MORRISON200798} and Lee \& Narayanan \cite{min2005} datasets include recordings from call-centers, with interactions involving a human agent and a machine agent, respectively. Both datasets primarily contain neutral speech, resulting in highly unbalanced data. Additionally, unlike previous datasets, those are restricted to public use. SUSAS dataset\cite{hansen1997getting} contains two real-life sets: isolated words from aircraft communications and doctor-patient interviews, but only the former is commercially available.

For Spanish, we found only two datasets in the literature: MOUD \cite{rosas2013multimodal} and CMU-MOSEAS \cite{zadeh2020}. Both datasets are publicly available; however, they primarily consist of monologue clips sourced from YouTube videos, recorded in staged settings. Additionally, the original clips in CMU-MOSEAS are not disclosed, with only high-level features provided. The MOUD dataset presents another limitation, with the majority of speakers being female (84 out of 105), while the CMU-MOSEAS dataset does not specify gender distribution, making it challenging to develop gender-fair SER models.

% -------------------------------------------------------------------------
%  SER permormance
% -------------------------------------------------------------------------
\subsection{Performance of SER models}
\label{sec:serperformance}

Research on SER using Spanish datasets is limited. For this reason, we reviewed prior studies that utilized Spanish datasets, including non-spontaneous ones, focusing on adult speech. Our analysis identified twelve Spanish datasets. Table \ref{tab:databases_spanish} summarizes key details: expression authenticity (acted, elicited, or spontaneous), recording environment (RE), number of samples (N), number of speakers (Spk) with gender distribution (males/female, M/F), emotion model (with the number of categories for discrete models), and free use. For the four multilingual datasets (marked with an asterisk), only the Spanish partition is detailed. Additionally, we report the classification performance of SER models from the literature for each Spanish datasets, using two evaluation metrics: Weighted Accuracy (WA) and Unweighted Accuracy (UA). WA, or simply ``accuracy", measures the overall proportion of correct predictions but can be biased toward majority classes in imbalanced datasets. UA, commonly used in SER studies, provides a fairer evaluation by averaging accuracy across all classes. The results indicate if a speaker-dependent (SD) or speaker-independent (SI) approach was used, i.e. whether same or different speakers were used for training and testing (if reported in the source articles). Multilingual and cross-lingual models are excluded from the table, as they pose greater challenges, and monolingual models typically achieve better performance when trained on comparable sample sizes \cite{atmaja2023multilingual}.

\begin{table*}[!hb]
    \caption{Datsets for SER in Spanish and corresponding results for SER models. An asterisk (*) denotes multilingual datasets.}
    \label{tab:databases_spanish}
    \centering
    \begin{tabular}{llllllll} \hline
        \multirow{2}{*}{Dataset} & \multirow{2}{*}{RE} & \multirow{2}{*}{N} & \multirow{2}{*}{Spk (M/F)} & \multirow{2}{*}{Emotion} & \multirow{2}{*}{Free} & \multicolumn{2}{l}{Models results} \\ \cline{7-8} 
         &  &  &  &  &  & UA (\%) & WA (\%) \\ \hline
        % ----------------------------------------------------------
        \textit{Acted} &  &  &  &  &  &  & \\ \hline
        % ----------------------------------------------------------
        Iriondo et al. \cite{iriondo2000validation}  & Lab & 336 & 8 (4/4) & Discrete (7) & No & - & - \\  
        Spanish Emotional Speech  \cite{montero1999analysis} & Lab & 1288 & 1 (1/0) & Discrete (4) & No & - & - \\ 
        Martínez \& Cruz \cite{martinez2005emotion} & Lab & 380 & $>$15 (-/-) & Discrete (5) & No & - & 55 \cite{martinez2005emotion} \\ 
        Emotional Mexican Spanish speech \cite{caballero2013recognition}   & Lab & 240 & 6 (3/3) & Discrete (4) & No & $\geq95$ \cite{caballero2013recognition} & - \\ 
        Spanish Expressive Voices \cite{barra2008spanish}  & Lab & 3890 & 2 (1/1) & Discrete (7) & No & - & 95 \cite{barra2008spanish} \\ 
        RekEmozio * \cite{lopez2007validating}  & Lab & 2618 & 10 (5/5) & Discrete (7) & No & - & SI: 74.82 \cite{arruti2014feature}\\ 
        Mexican Emotional Speech Database \cite{duville2021mexican}  & Lab & 3456 & 8 (4/4) & Discrete (6) & Yes & 93.90 / 89.49 M/F  \cite{duville2021mexican} & - \\  
        \begin{tabular}[t]{@{}l@{}}Emotional speech synthesis \cite{elra} \\  (from INTERFACE * \cite{hozjan2002interface})  \end{tabular} & Lab & 5520 & 2 (1/1) & Discrete (7) & Yes & 89.4 / 90.9 M/F \cite{duville2021mexican2}  & \begin{tabular}[t]{@{}l@{}}91.16 \cite{kerkeni2019automatic}, 91.51 \\ \cite{gustavo2020}, 94.01 \cite{kerkeni2019automatic2} \end{tabular} \\ 
        EmoFilm * \cite{parada2018} & Staged & 342 & 57 (33/24) & Discrete (5) & Yes & SI: 69.15 \cite{atmaja2023multilingual} & \begin{tabular}[t]{@{}l@{}} SI: 70.76 \cite{atmaja2023multilingual} \\ 
        85.0 \cite{costantini2022} \\  97.7 / 91.8 M/F \cite{costantini2022} \end{tabular}  \\  \hline
        % ----------------------------------------------------------
        \textit{Elicited} &  &  &  &  & &  &  \\ \hline
        % ----------------------------------------------------------
        \begin{tabular}[t]{@{}l@{}}EmoSpanishDB, \\ EmoMatchSpanishDB \cite{garcia2023emomatchspanishdb} \end{tabular} & Lab & \begin{tabular}[t]{@{}l@{}}3550, \\ 2020 \end{tabular} & 50 (30/20) & Discrete (7) & Yes & \begin{tabular}[t]{@{}l@{}}SD: 56.3, SI: 42.1 \cite{garcia2023emomatchspanishdb} \\ SD: 65.0, SI: 64.2 \cite{garcia2023emomatchspanishdb} \end{tabular} & - \\ \hline 
        % ----------------------------------------------------------
        \textit{Spontaneous} &  &  &  &  & &  &  \\ \hline
        % ----------------------------------------------------------
        MOUD \cite{rosas2013multimodal} & Staged & 550 & 105 (21/84) & \begin{tabular}[t]{@{}l@{}} Discrete (3) \\ from valence \end{tabular} & Yes & - & \begin{tabular}[t]{@{}l@{}} 46.75 (positive \\ vs negative) \cite{rosas2013multimodal} \end{tabular} \\  
        CMU-MOSEAS * \cite{zadeh2020} & Staged & 10000 & 341 (-/-) & Discrete (6) & Yes & - & - \\ \hline 
    \end{tabular}
\end{table*}

SER performance is conditioned by the type of data used. For acted datasets, the table shows WA and UA values equal to or exceeding 85\% (e.g. \cite{caballero2013recognition, barra2008spanish, duville2021mexican, duville2021mexican2, kerkeni2019automatic, gustavo2020, kerkeni2019automatic2, costantini2022}). However, performance drops to 70 to 75\% \cite{atmaja2023multilingual, arruti2014feature} when SI models are used, an approach which is known to yield lower results compared to SD models \cite{madanian2023}. An exception is seen in \cite{martinez2005emotion}, where an accuracy of 55\% was achieved for the classification of five emotions. The model was trained on actors simulating emotions while reading sentences and tested on movie clips, leading to a domain mismatch that likely caused the significant performance drop. Lower results on SI models also occur in the elicited EmoSpanishDB dataset \cite{garcia2023emomatchspanishdb}, where seven emotions where classified with UA scores of 56.3\% and 42.1\% for SD and SI, respectively. A refined version of the dataset, EmoMatchSpanishDB, improved performance by removing mislabeled samples that did not align with the intended elicited emotions, achieving UA scores of 65.0\% (SD) and 64.2\% (SI). Finally, the spontaneous dataset shows the lowest performance. In \cite{rosas2013multimodal}, a classification accuracy of 46.75\% was achieved for differentiating between positive and negative samples, highlighting the challenges of spontaneous emotional data in SER. 

Regarding the techniques used to create the SER models, a noteworthy paradigm shift involves leveraging large pre-trained models in emotion recognition tasks. The study in \cite{atmaja2023multilingual} used the pre-trained wav2vec2-large-robust model fine-tuned on MSP-Podcasts dataset from \cite{wagner2023} to extract speech embeddings. These embeddings were then fed into a support vector machine, obtaining 69.15\% unweighted accuracy for five emotion categories on the EmoFilm dataset. To our knowledge, this is the only previous study using pre-trained models for monolingual SER models in Spanish. Nevertheless, an increasing number of recent publications have focused on the use of pre-trained models with English datasets for emotion categories and less frequently for dimensions. 

As for studies focused on discrete emotion models, a very recent study \cite{wagner2023} summarized the state-of-the-art results for 4-class emotion classification in the widely used IEMOCAP dataset, including studies using the pre-trained wav2vec 2.0 and HuBERT models. Comparing the cross-validation results, the UA values range from 60.0\% to 74.3\%, while the weighted average recall values vary from 62.6\% to 79.6\%, both top models. Overall, HuBERT surpassed the performance of the wav2vec 2.0 models. Other works have explored several pre-trained models and emotional datasets. In \cite{phukan2023}, they used eight pre-trained models (including wav2vec 2.0) and four datasets to predict 6-7 emotions. The results showed that pre-trained models for speaker recognition (x-vector and ECAPA) achieved the highest scores, possibly due to their learned ability to identify unique features within an individual's speech. Subsequently, UniSpeech-SAT achieved the best results, a model pre-trained using multitask learning, including the speaker identity. In another study \cite{atmaja2022}, the authors compared nineteen pre-trained models (including wav2vec 2.0, HuBERT, UniSpeech-SAT and wavLM) and five datasets to predict 4-6 emotion categories using a SI approach. They found that the best scores were achieved with WavLM, UniSpeech-SAT, and HuBERT, all three in the large version.

As for studies on continuous emotion models, \cite{srinivasan2022} used a multimodal model (audio + text) as a teacher to fine-tune HuBERT embeddings to predict valence, arousal and dominance on MSP-Podcast dataset \cite{lotfian2017}. They obtained state-of-the-art Concordance Correlation Coefficient (CCC) values of 0.757, 0.627 and 0.671 for arousal, valence and dominance, respectively. The previous state-of-the-art performance for valence was 0.377 \cite{li2021contrastive}, so 0.627 represents a substantial improvement. They also replicated the results for the IEMOCAP dataset, achieving again state-of-the-art CCC results for valence (0.667), arousal (0.582) and dominance (0.545). In \cite{wagner2023}, the authors conducted an exhaustive analysis using different variants of the pre-trained wav2vec 2.0 and HuBERT models for valence, arousal and dominance prediction on MSP-Podcast. They obtained the best result in the literature for valence prediction using only audio, with a wav2vec2-large-robust model that achieved a CCC of 0.638. Notably, their results showed that data used for pre-training the models and the fine-tuning of the transformer layers had a strong influence on valence prediction. Both factors played a role in shaping the models' ability to implicitly incorporate linguistic information embedded in the audio signal. This, in turn, accounts for their success in valence prediction, achieving similar performance to multimodal models that integrate explicit textual information. 

% ------------------------------------------------------------------------
%  Fairness evaluation
% ------------------------------------------------------------------------
\subsection{Fairness evaluation}

Finally, research on evaluating model fairness is limited, particularly in the context of pre-trained models \cite{gorrostieta19_interspeech}. Some previous works have investigated gender-based fairness in Spanish datasets, revealing differences in model performance between females and males. In \cite{duville2021mexican2}, the authors reported a female accuracy of 90.9\% and a male accuracy of 89.4\%, yet the dataset only included one male and one female speaker. Conversely, \cite{duville2021mexican} and \cite{costantini2022} showed a different pattern, with males achieving higher accuracy than females. Specifically, \cite{duville2021mexican} presented 89.49\% for females and 93.90\% for males (with 4 female and 4 male speakers), while \cite{costantini2022} obtained 91.8\% for females and 97.7\% for males using pre-trained models (with 33 male and 24 female speakers). The same trend has been found in some works on SER for English data. In \cite{gorrostieta19_interspeech}, they studied the fairness of SER systems using pre-trained models and observed a reduction of 0.234 in CCC for arousal among females as opposed to males on MSP-Podcast. In \cite{wagner2023}, they found that pre-trained models tend to exhibit greater fairness in predicting arousal and dominance than in valence. Notably, the majority of models showed higher CCC for females than for males for valence. Nevertheless, overall the speech representations obtained with pre-trained models seem to be invariant to domain, speaker, and gender.  Additionally, the authors in \cite{wagner2023} also explored fairness across individual speakers and found that different pre-trained models show overall consensus on categorizing speakers as `good' or `bad', obtaining lower CCC values for some individuals in the latter group. These findings underline the importance of incorporating fairness assessment in future research.

% ##########################################################################################
% REAL-LIFE SCENARIO: EMOVOME dataset
% ##########################################################################################
\section{Emotional Voice Messages (EMOVOME)}
\label{sec:EMOVOME}

\begin{figure*}[!hb]
    \centering
    \includegraphics[width=0.75\linewidth]{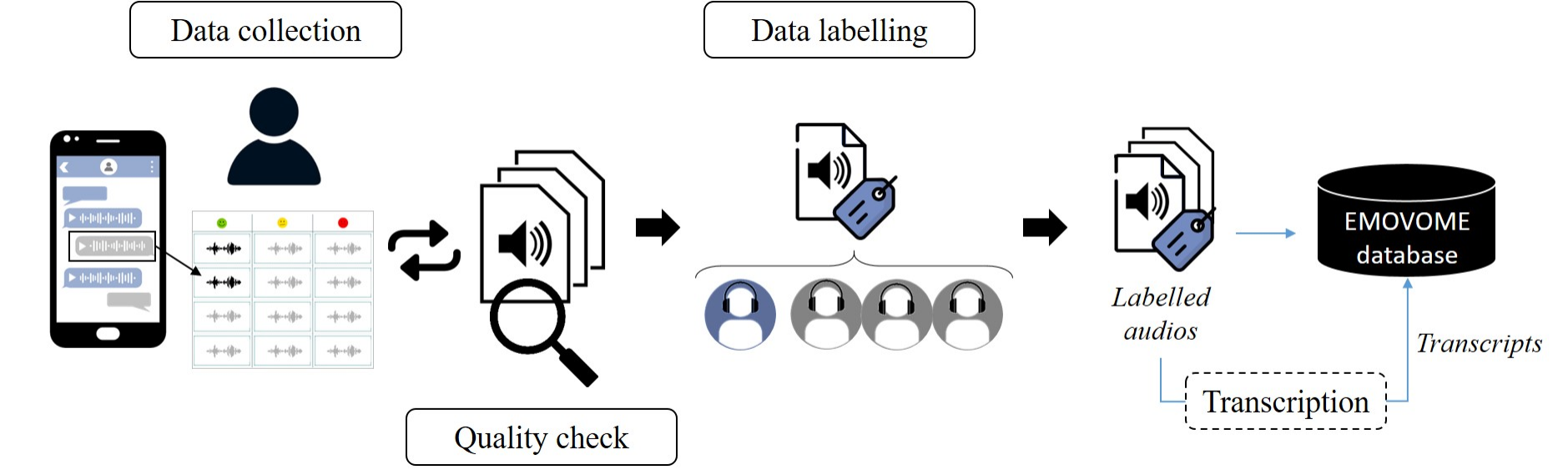}
    \caption{Overview of the methodology.}
    \label{fig:overview}
\end{figure*}

The Emotional Voice Messages (EMOVOME) dataset was created from scratch for this study is the first collection of spontaneous emotions from real voice messages. Figure \ref{fig:overview} presents a diagram illustrating the steps followed to create the EMOVOME dataset, detailed in the following subsections. The EMOVOME dataset is publicly available at the following Zenodo repository \url{https://zenodo.org/records/10694370}. Access to the raw audio files requires completing and signing an agreement form to ensure the speakers' privacy.

% ------------------------------------------------------------------------------------------
% Participants
% ------------------------------------------------------------------------------------------
\subsection{Participants}

The dataset includes 100 Spanish speakers (50\% females, 50\% males) equally distributed across four age ranges (18-25, 26-35, 36-45, and 46-55 years old), with no self-reported speech disorder, and who regularly send audio messages to their contacts. Additional demographic information can be found in the Appendix. Participants were sourced via a recruitment agency, which identified suitable individuals from their dataset based on the specified criteria and invited them to participate in the study for a monetary compensation of 25€. All methods and experimental protocols were performed in accordance with the regulations of the local ethics committee of the Universitat Politècnica de València (No. P02\_04\_06\_20).

% ------------------------------------------------------------------------------------------
% Data collection
% ------------------------------------------------------------------------------------------
\subsection{Data collection}

The sample collection was conducted through a web-based application designed for the study. Participants used their computer to complete the study, following on-screen instructions. Initially, they read the study protocol, provided informed consent, and completed a sociodemographic questionnaire. They then recorded an audio reading the short text in the Appendix. They also completed the NEO Five Factor Inventory (NEO-FFI) \cite{costa1989neo}, a 60-item questionnaire assessing personality traits. Next, participants were asked to upload 12 voice messages they had previously sent to other contacts, with a third of them being positive, neutral, and negative to ensure a balanced sample in terms of valence. Valence was chosen over specific emotion categories to simplify the task, as obtaining multiple samples for each distinct emotion can be challenging.  Platform instructions clarified the concepts of positive, neutral, and negative valence, providing examples for each. Voice messages were encoded using the Ogg Vorbis format.

% ------------------------------------------------------------------------------------------
% Quality check
% ------------------------------------------------------------------------------------------
\subsection{Quality check}

Voice messages were generated in real-life settings before participants were recruited, minimizing any bias associated with laboratory and staged environments. Recordings took place in participants' natural surroundings, ranging from quiet rooms at home to busy streets. To ensure high quality data, all recordings underwent a manual screening process. Audio files with critical background noise conditions (such as microphone malfunctions, background music or TV sounds) were identified and rejected. Participants were instructed to upload low-noise recordings, so if an audio was not considered suitable for the study, they were asked to send a new audio. We also excluded recordings that seemed to be recorded specifically for the study, as some participants may have attempted to fraudulently create these recordings after recruitment, despite instructions to upload voice messages from prior conversations. As a result, we received 1605 audio recordings, but 574 were not included in EMOVOME for this reason, resulting in 1031 audios. Finally, 32 audios exceeding 60 seconds were rejected to obtain a homogeneous sample concerning duration. Therefore, a total of 999 audios were considered in the final dataset (total duration of 293min), plus 100 audios reading the text. 

% ------------------------------------------------------------------------------------------
% Data labeling
% ------------------------------------------------------------------------------------------
\subsection{Data labeling}

The emotional content of the audios was labeled along two dimensions: valence, i.e., the degree to which an emotion is perceived as positive or negative; and arousal, i.e., how strongly the emotion is felt. Three non-experts and two expert evaluators were recruited for the task. The latter consisted of two clinical psychologists who rated half of the audios each. They are considered experts due to their professional training in recognizing and interpreting emotions. The demographic information of the annotators is indicated in the Appendix. Using the Self-Assessment Manikin (SAM) procedure \cite{bradley1994measuring}, the evaluators rated each audio's valence and arousal on a 5-point pictorial scale. Additionally, the experts provided an extra label for 7 emotion categories: the six basic emotions (happiness, disgust, anger, surprise, fear and sadness) defined by Ekman \cite{ekman1999} plus a neutral category. 

The valence and arousal ratings of the non-experts and expert evaluators present a V-shaped relation, that is, arousal increases with positive or negative valence (see the Appendix). To compare the ratings among evaluators, three categories for valence (positive, neutral and negative) and three for arousal (high, neutral and low) were defined by grouping negative and positive scores in the scale [-2, 2] and keeping 0 as neutral. The agreement between evaluators is presented in Table \ref{tab:agreement} using Cohen's kappa score, for both valence (V) and arousal (A). For valence, the kappa score goes from 0.614 to 0.754, indicating substantial agreement among annotators. In contrast, the agreement decreased considerably for the arousal dimension, with kappa values ranging from 0.075 to a maximum of 0.407. These results suggest valence recognition was easier for raters than arousal recognition. 

\begin{table}[!ht]
\caption{Pair-wise Cohen Kappa scores for non-experts (NE) and experts (E) pairs, for both valence (V) and arousal (A)}
\label{tab:agreement}
\centering
\begin{tabular}{|c|c|c|c|c|} \hline
            & NE 1 & NE 2 & NE 3 & E \\ \hline
            NE 1 & - &  \begin{tabular}[c]{@{}l@{}}V = 0.700\\ A = 0.142\end{tabular} & \begin{tabular}[c]{@{}l@{}}V = 0.754\\ A = 0.407\end{tabular} & \begin{tabular}[c]{@{}l@{}}V = 0.614\\ A = 0.275\end{tabular} \\ \hline
            NE 2 & \begin{tabular}[c]{@{}l@{}}V = 0.700\\ A = 0.142\end{tabular} & - & \begin{tabular}[c]{@{}l@{}}V = 0.732\\ A = 0.235\end{tabular} & \begin{tabular}[c]{@{}l@{}}V = 0.646\\ A = 0.075\end{tabular} \\ \hline
            NE 3 & \begin{tabular}[c]{@{}l@{}}V = 0.754\\ A = 0.407\end{tabular} & \begin{tabular}[c]{@{}l@{}}V = 0.732\\ A = 0.235\end{tabular} & - & \begin{tabular}[c]{@{}l@{}}V = 0.649\\ A = 0.241\end{tabular} \\ \hline
            E & \begin{tabular}[c]{@{}l@{}}V = 0.614\\ A = 0.275\end{tabular} & \begin{tabular}[c]{@{}l@{}}V = 0.646\\ A = 0.075\end{tabular} & \begin{tabular}[c]{@{}l@{}}V = 0.649\\ A = 0.241\end{tabular} & - \\ \hline
        \end{tabular}
\end{table}

\begin{figure}[!ht]
\centering
\includegraphics[width=0.47\textwidth]{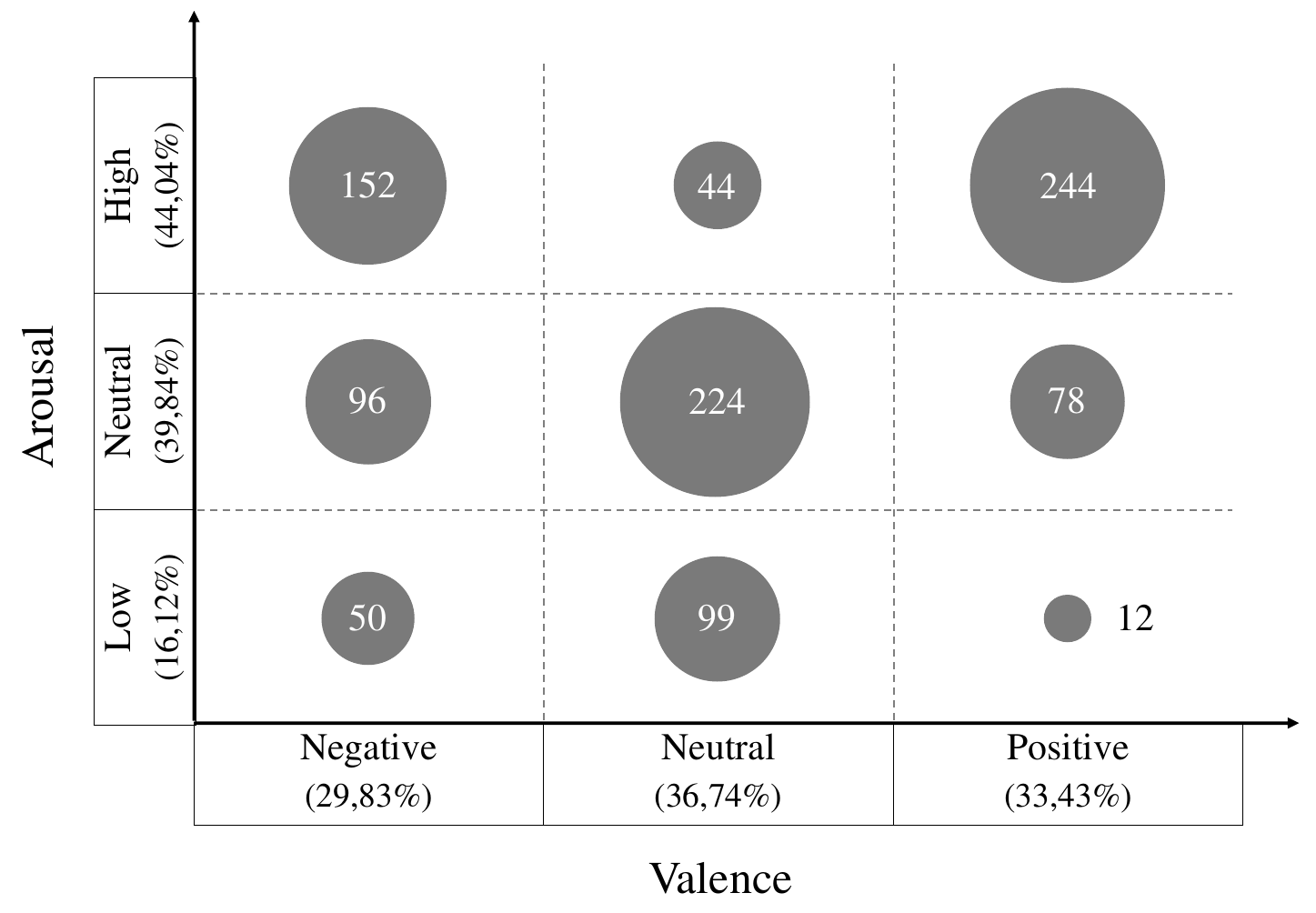}
\caption[Distribution of the nine clusters defined from the labels]{Distribution of audio samples based on their arousal and valence labels. The circle area is proportional to the number of samples in each group.} 
\label{fig:cluster_distribution}
\end{figure}

The non-experts and expert evaluations were combined to obtain a final audio label in terms of valence and arousal using the mode, i.e. the most frequent category among the four assessments. In instances of a two-way tie, we prioritized the expert's ratings as the final label for the audio. The distribution of the audio labels is shown in Figure \ref{fig:cluster_distribution}. Concerning valence, it is roughly balanced, with neutral being the majority class ($\sim$37\%) and negative the minority class ($\sim$30\%). Regarding arousal, the distribution is clearly unbalanced, with $\sim$44\% high, $\sim$40\% neutral and $\sim$16\% low arousal. Considering the combination of valence and arousal labels, the majority clusters are [positive valence, high arousal], [neutral valence, neutral arousal] and [negative valence, high arousal], with 244, 224 and 152 audios, respectively. The remaining groups have less than 100 audios, with a minimum of 12 audios in the cluster [positive valence, low arousal].

\begin{figure}[!ht]
\centering
\includegraphics[width=0.45\textwidth]{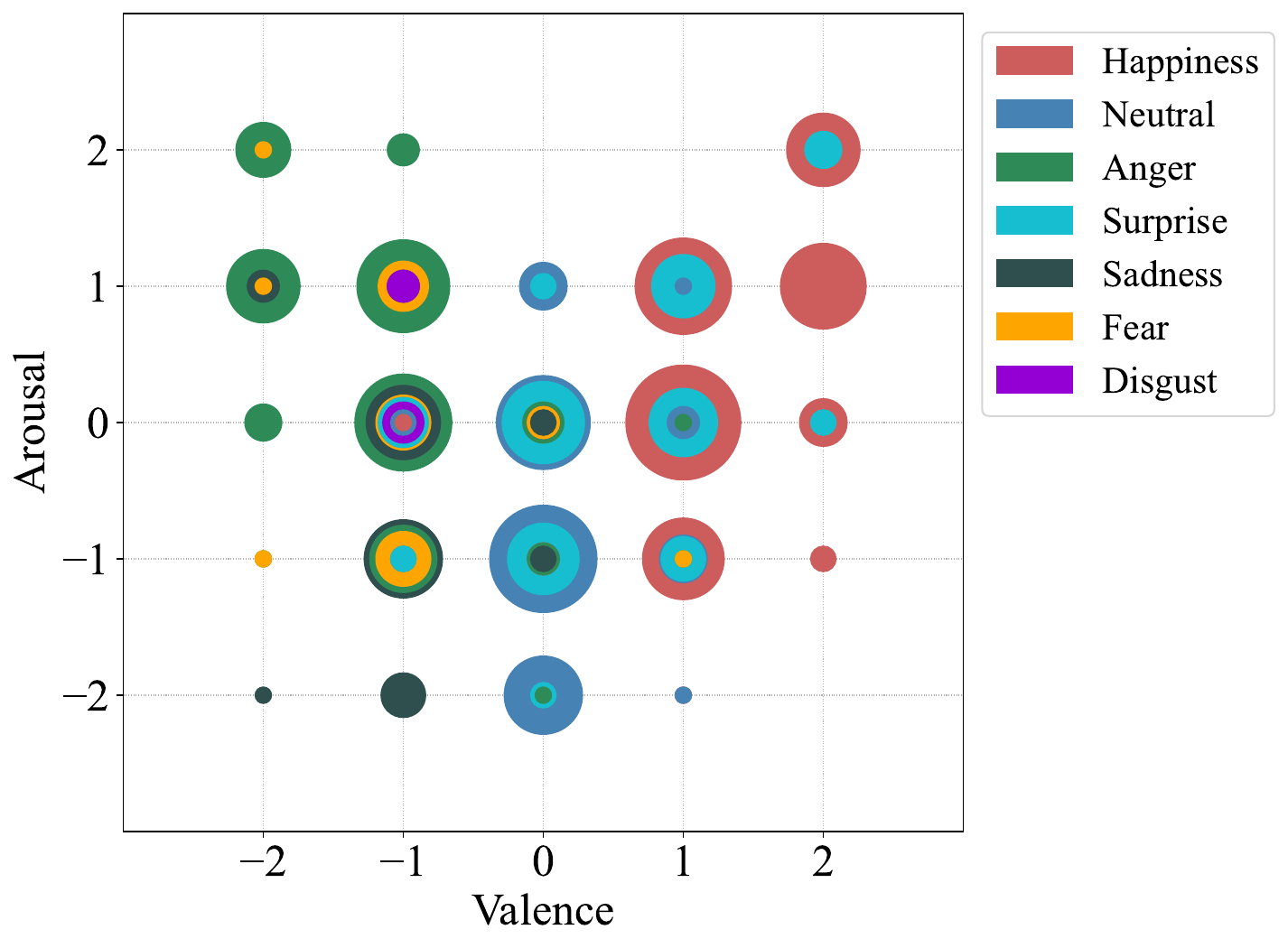}
\caption{Valence and arousal ratings for each emotion. The diameter represents the number of samples using a logarithmic scale}
\label{fig:categories_dist}
\end{figure}

Additionally, the expert annotators provided an extra label corresponding to 7 categories of emotions. The classes are distributed as follows: happiness (342), disgust (8), anger (199), surprise (118), fear (35), sadness (72) and neutral (225). Figure \ref{fig:categories_dist} shows the correspondence between categories and the valence and arousal dimensions provided by the experts. In this work, only the four most frequent emotions were used for classification, i.e., happiness, anger, neutral and surprise. 

% ------------------------------------------------------------------------------------------
% Transcription
% ------------------------------------------------------------------------------------------
\subsection{Transcription}

In addition to the speech files, audio transcriptions have been included in the dataset to offer an extra modality for analysis. The voice messages were transcribed using Amazon Transcribe and subsequently manually corrected to ensure accuracy and reliability. 

% ##########################################################################################
% Reference datasets: IEMOCAP and RAVDESS
% ##########################################################################################
\section{Reference datasets: IEMOCAP and RAVDESS}

In order to compare the novel natural dataset EMOVOME with the current state of the art, we selected two of the most used datasets: IEMOCAP and RAVDESS. A comparison of the main features is presented in Table \ref{tab:databases}.

\begin{table*}[!hb]
\caption{Comparison of the emotion datasets used in this work: EMOVOME, IEMOCAP and RAVDESS.}
\label{tab:databases}
    \centering
    \begin{tabular}{lllllllllll} \hline
        \multirow{2}{*}{Dataset} & \multirow{2}{*}{\begin{tabular}[c]{@{}l@{}}Expression \\ authenticity \end{tabular} } & \multirow{2}{*}{\begin{tabular}[c]{@{}l@{}}Recording \\ environment \end{tabular} } & \multirow{2}{*}{Language} & \multirow{2}{*}{Emotions} & \multirow{2}{*}{Raters} & \multirow{2}{*}{Samples} & \multirow{2}{*}{\begin{tabular}[c]{@{}l@{}}Speakers \\ (Male/Female) \end{tabular} } & \multirow{2}{*}{\begin{tabular}[c]{@{}l@{}}Samples/ \\ speaker \end{tabular}} & \multicolumn{2}{l}{Sample duration} \\ \cline{10-11} & & & & & & & & &  Mean & Range \\ \hline
        EMOVOME   & Natural & Real-life  & Spanish & \begin{tabular}[t]{@{}l@{}}Continuous  \\ Discrete (7) \end{tabular}   & 4 & 999       & 100 (50/50)   & 10 & 17.59s &  1-60s  \\ 
        IEMOCAP  & \begin{tabular}[t]{@{}l@{}}Elicited \\ and acted \end{tabular} & Lab & English & \begin{tabular}[t]{@{}l@{}}Continuous  \\ Discrete (9) \end{tabular} & 319 & 10039    & 10 (5/5) & 1004 & 4.59s & 1-35s  \\ 
        RAVDESS  & Acted & Lab  & English & Discrete (8)  & - & 1440      & 24 (12/12) & 60  & 3.70s &  3-5s  \\ \hline
    \end{tabular}
\end{table*}

% ------------------------------------------------------------------------------------------
% IEMOCAP dataset
% ------------------------------------------------------------------------------------------

IEMOCAP \cite{busso2008iemocap} stands out as a benchmark dataset for studying emotional expression and communication. It contains dyadic interactions of 10 English-speaking actors engaged in scripted and improvised dialogues designed to elicit different emotions. In this work, only the audio files corresponding to the segmentation of the conversations into utterances were used to create the SER models. This set includes 10039 utterances annotated into 9 emotion categories and 3 dimensions. To align the classification results with the EMOVOME dataset, valence and arousal scores were stratified into three categories. Samples scoring below 2.5 were deemed negative/low, those between 2.5 and 3.5 were neutral, and those above 3.5 were positive/high. To facilitate comparison with the literature, only the four most frequent emotions were used for the classification into categories, i.e., angry, neutral, sad, happy and excited (the latter two were merged, following previous studies). 

% ------------------------------------------------------------------------------------------
% RAVDESS dataset
% ------------------------------------------------------------------------------------------

RAVDESS \cite{livingstone2018} is a multimodal dataset with speech and song recordings expressing different emotions. In this study, only the voice data was used, where 24 actors recorded two phrases with normal and strong emotional intensity for each of eight emotions (excluding neutral), totaling 1,440 samples. To compare the results with EMOVOME, the emotion labels were transformed from the discrete model to the dimensional model, as also studied in \cite{RAVDESS2020dimensions}. The emotion categories were converted into valence and arousal dimensions following their distribution in Russell's circumplex model of affect \cite{russell1980circumplex}. For valence, the labels were: negative (sad, angry, fearful, disgust), neutral (neutral, surprised) and positive (happy, calm). For arousal, the labels were: low (calm, sad), neutral (neutral, disgust) and high (happy, angry, fearful, surprised).

% ##########################################################################################
% Methods
% ##########################################################################################
\section{Methods}

Three approaches were implemented to create the SER models. First,a baseline was established using a standard feature set with classical machine learning algorithms (Section \ref{methods:baseline}). Second, pre-trained models were used as feature extractors, followed by a linear layer for classification (Section \ref{methods:pretrained}). Finally, a combination of pre-trained models and standard features was analyzed (Section \ref{methods:pretrained_egemaps}). The Python implementation is available on the following GitHub repository: \url{https://github.com/LuciaGomZa/SER_EMOVOME.git}.

For evaluation, the datasets were split into 80\% for training and 20\% for testing using a speaker-independent approach. To facilitate reproducibility and comparison of results, for testing we used: a proposed set of 20 participants (50\% female) available in the EMOVOME Zenodo repository indicated in Section \ref{sec:EMOVOME}; session 5 for IEMOCAP; and ``fold 0" proposed in \cite{jimenez2022} for RAVDESS. Details of the label distribution are provided in the Appendix. Additionally, parameter tuning also employed a SI cross-validation scheme with 4 folds for IEMOCAP and 5 folds for EMOVOME and RAVDESS. All methods were evaluated using weighted and unweighted accuracy, described in Section \ref{sec:serperformance}.

% ------------------------------------------------------------------------------------------
% Baseline: eGeMAPS and machine learning
% ------------------------------------------------------------------------------------------
\subsection{Baseline: eGeMAPS and machine learning}
\label{methods:baseline}

The voice messages in EMOVOME were recorded in real-world conditions, using various devices -mostly smartphone microphones- which resulted in different sampling rates (94\% at 48 kHz and 6\% at 44.1 kHz). To standardize, EMOVOME audio samples were resampled to 44.1 kHz, while RAVDESS and IEMOCAP maintained 48 kHz and 16 kHz, respectively. Next, the extended Geneva Minimalistic Acoustic Parameter Set (eGeMAPS) \cite{eyben2015geneva} was extracted with the openSMILE toolkit \cite{eyben2010opensmile}. The use of this standard set of features facilitates the understanding and reproducibility of the results. The 88 features per audio were normalized by subtracting the mean and dividing by the standard deviation of the development samples. For feature selection, high-correlated features ($p>0.95$) were first eliminated using Pearson's correlation matrix, and a filter method was then used to select 25\%, 50\% or 75\% of the features based on the highest ANOVA F-values. Finally, Support Vector Machine (SVM) and K-Nearest Neighbours (KNN) models were fitted on the development set according to the cross-validation scheme indicated above. Hyperparameter tuning was performed for both models. For SVM, we changed the kernel (radial basis function, sigmoid), gamma (0.001, 0.01, 0.1, 1, 'auto', 'scale') and C (1, 10, 100, 1000). For KNN, we optimized the number of neighbors (1, 3, 5, 7), weights (uniform, distance) and metric (Euclidean, Manhattan, Minkowski). The chosen combination of features and hyperparameters was used to train a model on the entire development set, followed by an evaluation of the test set.

% ------------------------------------------------------------------------------------------
%Pre-trained model embeddings
% ------------------------------------------------------------------------------------------
\subsection{Pre-trained model embeddings}
\label{methods:pretrained}

We evaluated several pre-trained models with different architectures, pre-training methodologies and pre-training audio data. First, several variations of the widely used Wav2vec 2.0 model were selected. Considering that EMOVOME is in Spanish, and we wanted to compare it with other two datasets in English, we selected models pre-trained using a multilingual approach: facebook/wav2vec2-xls-r-300m (w2v2-xlsr-128) \cite{babu2021xls} pre-trained on 436k hours of audios in 128 languages, and facebook/wav2vec2-large-xlsr-53 (w2v2-xlsr-53) \cite{conneau21_interspeech}, pre-trained on 56k hours of audios in 53 languages. Both models included Spanish in the pre-training data. Despite prior research suggesting that fine-tuning models for automatic speech recognition (ASR) do not help with speech emotion recognition \cite{wang2021fine, pepino21_interspeech}, we decided to include a fine-tuned model for Spanish ASR. This decision aimed to explore whether this approach outperforms the utilization of a pre-trained model in a different language since \cite{wang2021fine} and \cite{pepino21_interspeech} used pre-trained models in the same language as the dataset tested (English). Therefore, we also used the fine-tuned version facebook/wav2vec2-large-xlsr-53-spanish (w2v2-xlsr-53-spa) \cite{conneau21_interspeech} to obtain the embeddings. We also included a model pre-trained using noisy audios, the model facebook/wav2vec2-large-robust (w2v2-L-robust) \cite{hsu2021robust}, as it may be useful for the EMOVOME dataset despite being in English. As a widely-used alternative to Wav2Vec2, we used a HuBERT model, particularly the large version facebook/hubert-large-ll60k (hubert-L) \cite{hsu2021hubert}. Additionally, recent studies \cite{phukan2023,pappagari2020,moine21_interspeech} have indicated that including information about the speaker is helpful for speech emotion recognition. Therefore, following these investigations, two more models were selected: Microsoft's UniSpeech-SAT-Large (unispeech-L) \cite{chen2022unispeech} and a Statistics Pooling Time Delay Neural Network to obtain x-vector embeddings \cite{snyder2018spoken}. The former is a model pre-trained using multitask learning, including also the speaker identity during training. The latter provides a speaker embedding learning during a speaker verification task. All the pre-trained models are available in HuggingFace. The x-vector also requires the Speechbrain toolkit \cite{speechbrain}. All of them require the input audio to be resampled to 16 kHz.

Previous investigations have adapted the architecture of the pre-trained models by adding a classification head and fully or partially fine-tuning the model. Nevertheless, this process requires high computational resources due to the large size of some models and the audio lengths (particularly in EMOVOME). For this reason, the embedding extraction process was implemented offline, saving the audio embeddings in independent files. For all pre-trained models except for the x-vectors, we extracted the last hidden state of the last transformer layer. As a result, we obtained a vector of dimensions $(X, 1024)$, where $X$ varies depending on the audio length. Following previous research \cite{wagner2023, wang2021fine}, we applied the average over the time dimension to obtain a 1024-dimension vector per audio sample. For x-vectors, the pre-trained model has a built-in statistics pooling layer that computes the mean and standard deviation of information from the last frame-level layer. These statistics are combined and input into a 512-dimensional hidden layer that is finally used to obtain the embeddings. As a result, the pre-trained model consistently outputs a 512-dimensional vector, independent of audio length, eliminating the need for extra aggregation strategies. Subsequently, we trained a neural network comprised solely of a linear layer, which took the embeddings as input and produced the output corresponding to the number of labels. This architecture, proven effective in previous literature \cite{wang2021fine}, simplifies the model while still capturing essential features in the merged embeddings. The model was trained during a maximum of 3000 epochs during cross-validation using Adam optimization, a learning rate of 0.001 and a batch size of 128. The early stopping callback was applied, set to monitor the validation loss with a patience of 50. For the selected hyperparameters, a final model was trained on the development set for a fixed number of epochs selected based on the cross-validation results and evaluated on the test. 

% ------------------------------------------------------------------------------------------
% Pre-trained model embeddings and eGeMAPS
% ------------------------------------------------------------------------------------------
\subsection{Pre-trained model embeddings and eGeMAPS}
\label{methods:pretrained_egemaps}

We integrated previous methods by combining speech embeddings with the eGeMAPS feature set. The 1024-dimensional embeddings (except for the x-vector, which is 512-dimensional) were concatenated with the 88 eGeMAPS features. The combined set underwent normalization before being fed into the neural network, which consists of a linear layer, following the procedure detailed in the previous section.

% ##########################################################################################
% Results
% ##########################################################################################
\section{Results}

\begin{figure*}[!hb] 
    \centering
    \includegraphics[width=\textwidth]{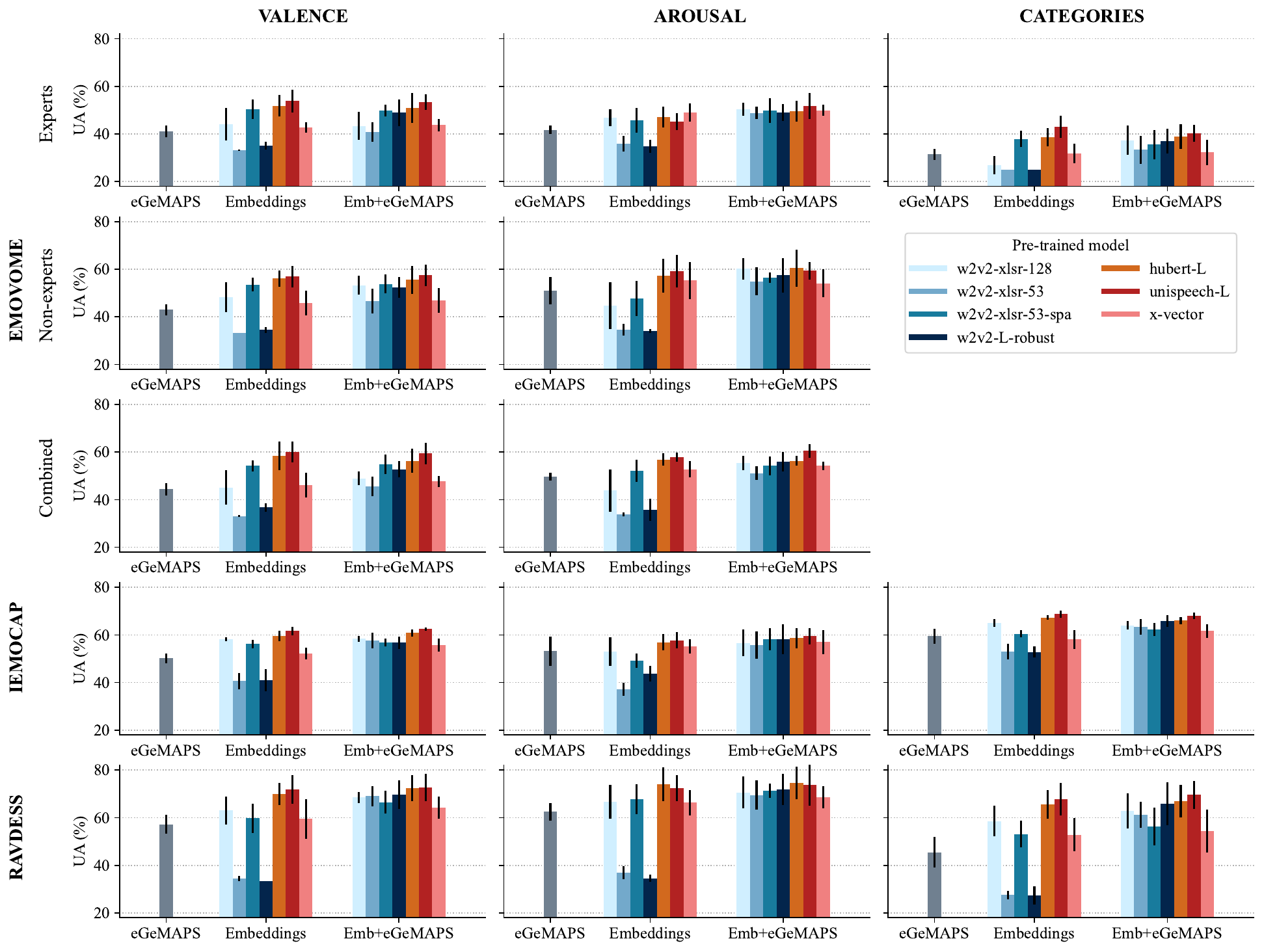}
    \caption[]{Cross-validation results for the three methods implemented (eGeMAPS, Embeddings and Emb+eGeMAPS) across the different datasets (EMOVOME, IEMOCAP and RAVDESS) and emotion labels (valence, arousal and categories of emotions).}
    \label{fig:CV_results}
\end{figure*}

The cross-validation results in Fig. \ref{fig:CV_results} show the performance across datasets and emotion labels. The UA of the optimal baseline model using eGeMAPS features and machine learning is indicated in gray. Next to this, results for models using embeddings are displayed, along with a third column for the method integrating embeddings and eGeMAPS features (Emb+eGeMAPS). The figure indicates the mean and standard deviation of the UA across the five folds (four for IEMOCAP).

Regarding the baseline eGeMAPS models, the lowest UA values are from the EMOVOME dataset, with 41-44\% for valence, 41-51\% for arousal and 31.42\% for four categories. IEMOCAP achieves 50.20\%, 53.21\% and 59.50\% for valence, arousal and categories, respectively. RAVDESS achieves the highest UA results for valence and arousal, with 57.19\% and 62.48\%, respectively. For the eight categories, it obtains 45.52\%. Overall, SVC was preferred to KNN in most cases. 

The embedding approach demonstrates a notable enhancement in the models' performance compared to the baseline method. Again, EMOVOME yields the lowest scores, with UA values in the range 54-60\% and 49-59\% for valence and arousal, respectively, and 43.30\% for the 4-way classification. IEMOCAP models have UA results of 61.73\%, 57.68\% and 68.79\% for valence, arousal and categories, respectively. RAVDESS again reaches top UA values, with 71.91\%, 74.09\% and 67.81\% for valence, arousal and categories, respectively. In general, unispeech-L consistently demonstrated superior performance compared to alternative options. Using pre-trained embeddings results in an enhancement of approximately 10\% in the UA across all combinations.

Finally, the Emb+eGeMAPS method obtains similar results to the previous approach. EMOVOME achieves similar values in valence and arousal (53-59\% and 52-60\% respectively), while the UA in 4-emotion classification decreases (40.22\%). A possible reason is that EMOVOME was recorded in-the-wild conditions. Therefore, eGeMAPS features may be affected by microphone quality and background noise \cite{eyben2015geneva}. Conversely, IEMOCAP and RAVDESS were recorded in a controlled environment, and the UA values increased around 1-3\% compared to the previous approach (except for the emotion categories with IEMOCAP, which slightly decreased). IEMOCAP achieves 62.48\%, 60.41\% and 68.04\% for valence, arousal and categories, respectively. RAVDESS obtains 72.70\%, 75.27\% and 69.58\% for valence, arousal and categories, respectively. Again, unispeech-L is the best option in most cases. However other pre-trained models, with lower results in the previous approach, greatly improved when combined with eGeMAPS features, particularly w2v2-xlsr-53 and w2v2-L-robust. 

\begin{figure}[!ht]
    \centering
    \includegraphics[width=0.95\linewidth]{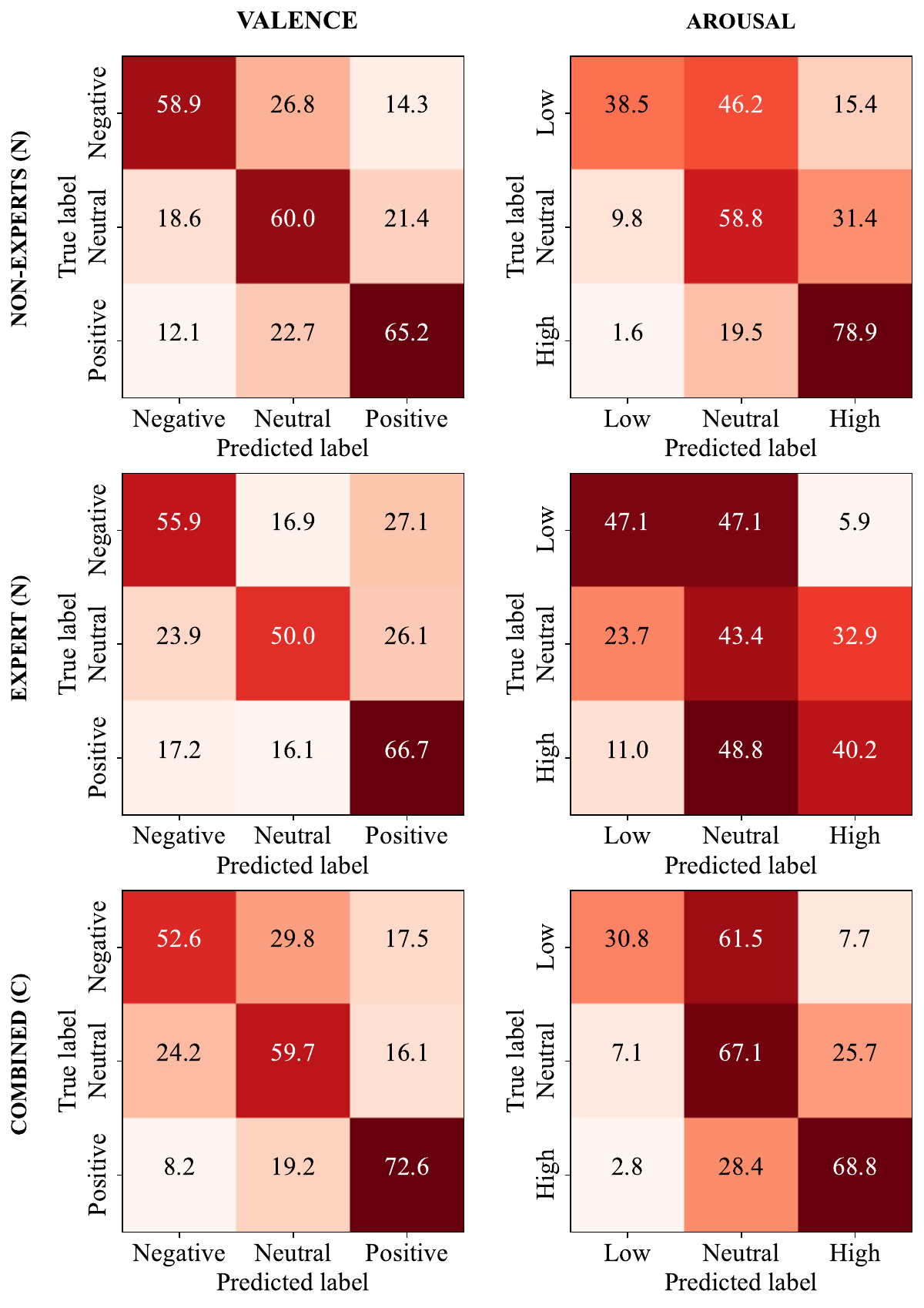}
    \caption{Confusion matrix for the test samples for valence and arousal prediction on EMOVOME, for the different raters.}
    \label{fig:cm_test_EMOVOME}
\end{figure}

Using the hyperparameters of the models obtaining the highest UA in cross-validation, new models were trained on the development set and evaluated on the test set, obtaining the results in Table \ref{tab:test_results_VA}. The model achieving the highest cross-validation results among the three implemented methods (eGeMAPS, Embeddings and Emb+eGeMAPS) is indicated. The test results follow the trends found in cross-validation, with RAVDESS outperforming IEMOCAP and EMOVOME across the two dimensions (73.54\% in valence and 71.94\% in arousal). IEMOCAP UA scores are higher than EMOVOME scores in arousal prediction (61.20\% vs. 43.57-58.73\% respectively), whereas, for valence prediction, IEMOCAP and EMOVOME achieve similar results (60.40\% for the former and 57.53-61.64\% for the latter). For EMOVOME, we also examined the differences between expert (E) and non-expert (N) annotations, as well as their combination (C). The confusion matrices are represented in Fig. \ref{fig:cm_test_EMOVOME}. 

\begin{table}[!ht]
\centering
\caption{Test results for 3-class valence and arousal prediction. ``PTM" specifies the pre-trained model. For EMOVOME, the rater is indicated: expert (E), non-expert (N), or combined (C).}
\label{tab:test_results_VA}
\begin{tabular}{lllll}
\hline
Dataset                        & Method      & PTM         & WA    & UA \\ \hline
\textit{Valence prediction}     &       &          &     &  \\ \hline
EMOVOME - E                      & Embeddings  & Unispeech-L & 59,38 & 57,53 \\
\textcolor{white}{EMOVOME} - N   & Emb+eGeMAPS & Unispeech-L & 61,46 & 61,36 \\
\textcolor{white}{EMOVOME} - C   & Embeddings  & Unispeech-L & 62,50 & 61,64 \\ \hline
IEMOCAP                         & Emb+eGeMAPS & Unispeech-L & 62,35 & 60,04 \\ 
RAVDESS                         & Emb+eGeMAPS & Unispeech-L & 77,33 & 73,54 \\ \hline
\textit{Arousal prediction}      &       &          &     &  \\ \hline
EMOVOME - E                      & Emb+eGeMAPS & Unispeech-L & 42,71 & 43,57 \\
\textcolor{white}{EMOVOME} - N   & Emb+eGeMAPS & Hubert-L & 70,83 & 58,73 \\ 
\textcolor{white}{EMOVOME} - C   & Emb+eGeMAPS & Unispeech-L & 65,62 & 55,57 \\ \hline
IEMOCAP                         & Emb+eGeMAPS & Unispeech-L & 74,38 & 61,20 \\ 
RAVDESS                         & Emb+eGeMAPS & Hubert-L & 77,00 & 71,94 \\ \hline
\end{tabular}
\end{table}

\begin{figure}[!ht]
    \centering
    \includegraphics[width=0.9\linewidth]{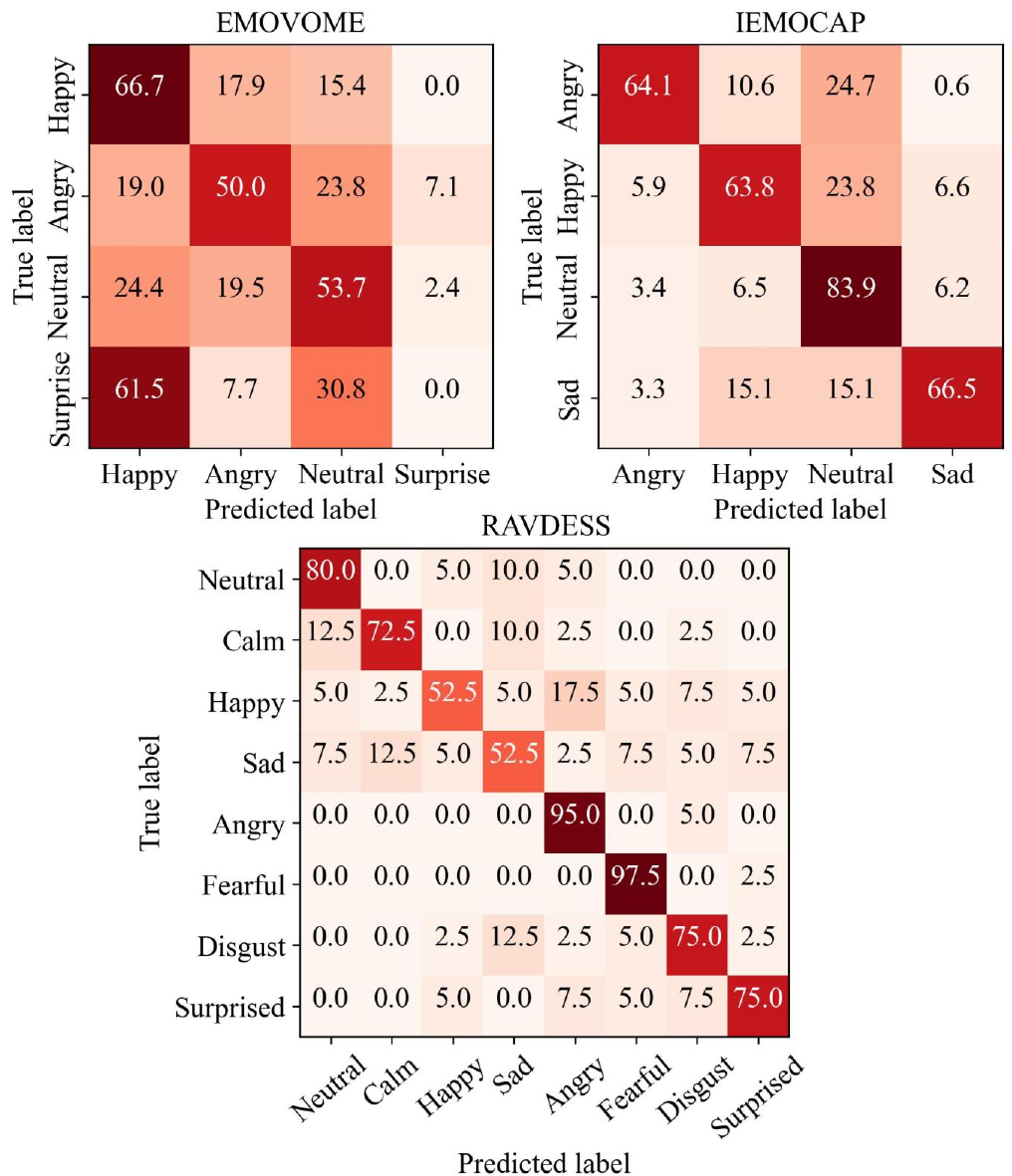}
    \caption{Confusion matrix for the test samples for emotion category prediction on EMOVOME, IEMOCAP and RAVDESS.}
    \label{fig:cm_test_categories}
\end{figure}

Considering the emotion categories, test results for the three datasets are presented in Table \ref{tab:test_results_categories}. It includes the model achieving the highest CV results among the three implemented methods. RAVDESS obtains the highest UA score (75.00\%), followed by IEMOCAP (69.58\%) and EMOVOME (42.58\%). To evaluate which emotions are misclassified, Fig. \ref{fig:cm_test_categories} shows the confusion matrix for each dataset.

\begin{table}[!ht]
\centering
\caption{Test results for emotion categories prediction. \\ ``PTM" specifies the pre-trained model employed.} 
\label{tab:test_results_categories}
\begin{tabular}{lllll}
\hline
Dataset    &  Method         & PTM           & WA & UA \\ \hline
EMOVOME - E & Embeddings     & Unispeech-L & 54,60 & 42,58 \\ 
IEMOCAP     & Embeddings     & Unispeech-L & 70,59 & 69,58 \\ 
RAVDESS     & Emb+eGeMAPS    & Unispeech-L & 74,67 & 75,00 \\ \hline
\end{tabular}
\end{table}

\begin{figure}[!ht]
    \centering
    \includegraphics[width=\linewidth]{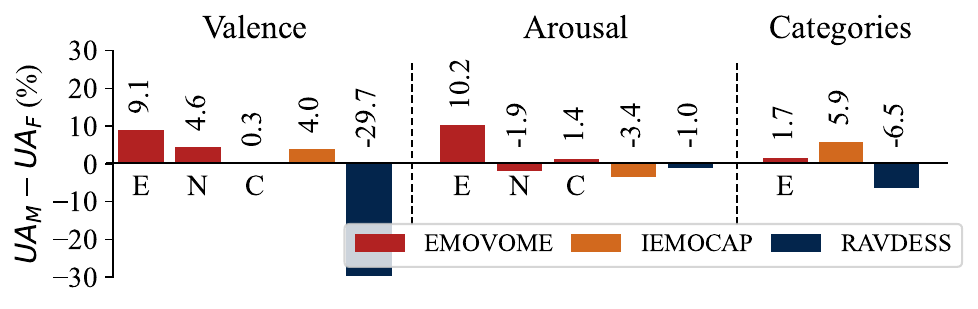}
    \caption{Evaluation of gender fairness for the three labels and datasets.}
    \label{fig:fairness_gender}
\end{figure}

Finally, we evaluated model fairness in terms of gender by calculating the difference between the UA for male speakers ($UA_M$) and the UA for female speakers ($UA_F$) on the test set, which is presented in Fig. \ref{fig:fairness_gender}. A positive difference means that the model had a better performance for male speakers. 

% ##########################################################################################
% Discussion
% ##########################################################################################
\section{Discussion}

In this study, we addressed the scarcity of spontaneous datasets from real-life environments by developing the EMOVOME dataset—the first public repository featuring genuine emotions from spontaneous conversations in real-life settings. This paper provides a detailed description of EMOVOME and extends access to the scientific community. We evaluated the performance of SER models trained with EMOVOME using various methodologies, and we compared the results with the well-known IEMOCAP dataset, which includes elicited speech, and RAVDESS, comprising acted recordings. Our comparison focused on predicting valence, arousal, and emotion categories. Finally, we also investigated the impact of annotators' labels and speakers' genders on model performance. UA was used for comparison since it is more suitable for unbalanced data.

% ------------------------------------------------------------------------------------------
% SER performance
% ------------------------------------------------------------------------------------------
\subsection{SER performance}

We implemented three methods to create the SER models: the baseline using eGeMAPS features and machine learning, the embeddings from pre-trained models and the integration of embeddings and eGeMAPS features. As shown in Fig. \ref{fig:CV_results}, overall, the embedding approach significantly improves model performance compared to the non-transformer baseline, leading to an approximately 10\% improvement in unweighted accuracy across all combinations. Unispeech-L consistently showcased better results than other pre-trained models (as shown in \cite{phukan2023}), closely followed by hubert-L (which also outperforms wav2vec2 models in previous studies \cite{wagner2023}). For the English datasets, in general, the models pre-trained on multiple languages performed worse than those trained on English-only data (as in \cite{wagner2023}). Regarding the Emb+eGeMAPS method, it shows similar cross-validation results to the embeddings for the EMOVOME dataset, and UA values slightly increase (max. 3\%) for IEMOCAP and RAVDESS in some cases. One potential explanation is that EMOVOME was recorded in natural, uncontrolled conditions, which might impact the reliability of eGeMAPS features, unlike the other two datasets that were conducted in a controlled environment. Unispeech-L is again the top choice in most cases, but notably, other pre-trained models that initially performed lower (e.g. w2v2-xlsr-53 and w2v2-L-robust) show significant improvement when combined with eGeMAPS features. 

Considering the test results, the prediction of emotion categories (see Table \ref{tab:test_results_categories}) follow the trend in cross-validation, where EMOVOME obtains the lowest evaluation metric (45.58\% UA), followed by IEMOCAP (69.58\% UA) and finally RAVDESS (75.00\% UA) (even though the first two classify four emotions and the later eight). Although IEMOCAP contains elicited speech, part of the data consists of actors performing scripted dialogues, which could lead us to expect higher classification results. However, IEMOCAP contains utterances whose text content is different throughout the dataset, i.e., it is text-independent (same as EMOVOME). Conversely, RAVDESS is text-dependent, using two fixed sentences. This might have led the models for RAVDESS to prioritize emotional variations over differences in semantic content. Consequently, there is a performance gap in IEMOCAP compared to the other acted datasets, as highlighted in \cite{fahad2021survey}. Nevertheless, our classification results (70.59\% WA, 69.58\% UA) are comparable to the state of the art for IEMOCAP, which is in the range from 60.0\% to 74.3\% UA \cite{wagner2023}, especially considering that we implemented a SI approach, unlike other previous studies. Additionally, we examined the misclassified emotions for each dataset in Fig. \ref{fig:cm_test_categories}. For EMOVOME, surprise is not correctly predicted in any case and is mainly confused with happy and neutral emotion, but this category is underrepresented in the data. For IEMOCAP, all emotions are mainly mistaken for the neutral category. For RAVDESS, most emotions are accurately classified ($>$72\%).

Regarding the test results for arousal and valence dimensions (see Table \ref{tab:test_results_VA}), the SER models for the EMOVOME dataset achieve UA values of 61.64\% for valence and 55.57\% for arousal, considering the combined label between expert and non-experts. Surprisingly, the IEMOCAP dataset obtains similar values to EMOVOME, 60.04\% for valence and 61.20\% for arousal. As for RAVDESS, the UA values are 73.54\% and 71.94\% for valence and arousal, respectively. Initially, one might have anticipated better outcomes in valence, given that earlier studies \cite{triantafyllopoulos2022, wagner2023} found that pre-trained models inherently capture linguistic information in the audio, aiding valence prediction. However, in our approach, we utilized pre-trained embeddings solely as feature extractors to obtain speech embeddings without fine-tuning the transformer layers for SER. This step has proven to be fundamental for the models to effectively learn the semantic content \cite{wagner2023}. Furthermore, the data used for pre-training significantly impacts the models' capacity to capture linguistic information, with the inclusion of multi-lingual data adding complexity to the task \cite{wagner2023}. These considerations might explain the relatively minor differences observed between valence and arousal predictions. In the case of RAVDESS, the use of fixed semantic content during recording prevented pre-trained models from leveraging text information for valence prediction. Furthermore, an important limitation in both IEMOCAP and RAVDESS is the label transformation applied to obtain the valence and arousal categories, as well as the unbalanced distribution of samples. 

In summary, we found that embeddings and Emb+eGeMAPS give similar results for the top-performing pre-trained model, i.e., Unispeech-L. The EMOVOME dataset achieves lower results than other non-natural datasets in the literature. Both RAVDESS and IEMOCAP outperform EMOVOME in emotion categories, and the former also obtains higher valence and arousal prediction results. However, EMOVOME and IEMOCAP exhibit more comparable results in valence and arousal, possibly owing to their text-independent nature (unlike RAVDESS) and the approach used to transform the original labels in IEMOCAP into valence and arousal categories.

% ------------------------------------------------------------------------------------------
% Impact of annotator labels on SER performance
% ------------------------------------------------------------------------------------------
\subsection{Impact of annotator labels on SER performance}

Emotion prediction lacks a definitive ground truth due to its inherently subjective nature. In this study, we addressed this by asking participants to upload audio samples with self-assessed emotional valence (negative, neutral, or positive). To refine the emotional labels, we also incorporated an external evaluation by both experts and non-experts. While research on the impact of annotators' demographics on SER is limited, existing studies suggests that biases can arise based on their gender, age or educational level \cite{ding2022}. We hypothesized that this bias may be more pronounced when evaluating natural datasets,such as EMOVOME, since they do not contain stereotypical emotion and thus can be more challenging to label. In this work, we explored the difference between expert (E) and non-expert (N) annotations, as well as their combination (C) (see Table \ref{tab:test_results_VA}). 

Surprisingly, SER models using the expert's labels achieved the lowest results for both valence and arousal. Non-experts got higher UA values in arousal prediction (58.73\%), compared to the combined labels (55.57\%) and the expert's labels (43.57\%). For valence, the combined label obtained the highest UA score (61.64\%), closely followed by the non-experts (61.36\%) and lastly the expert (57.53\%). It's essential to note that while clinical psychologists are experts in the task, emotions remain highly subjective and can be influenced by individual experiences \cite{ding2022}. In fact, the confusion matrix (see Fig. \ref{fig:cm_test_EMOVOME}) reveals discernible biases towards specific emotion categories in both valence and arousal results, potentially shaping what the SER models learn. For valence, the model trained on expert labels shows a bias towards positive valence. Conversely, the model trained on non-expert labels tends to misclassify neutral samples as either positive or negative. The combined label model mitigates the bias towards positive expert categories but increases misclassification for negative and neutral categories. In arousal, unlike valence, there's an unbalanced data distribution. Expert labeled 43\% of training data as neutral arousal, leading the model to often assign this category to test samples. Non-experts show a bias toward high arousal. Combining labels mitigates expert bias toward the neutral class and reduces non-expert bias toward the positive category. As a result, the model has fewer low arousal samples to train on, lowering accuracy in this category.

Overall, the annotators' biases may cause differences in UA scores to up to 4\% for valence and 15\% for arousal. The better performance of the models based on the non-experts and the combined labels may be due to the higher number of annotators included, which may reduce individual biases in their interpretation of emotions. However, this number is still limited compared to the 319 raters in IEMOCAP.

% ------------------------------------------------------------------------------------------
% Fairness evaluation
% ------------------------------------------------------------------------------------------
\subsection{Fairness evaluation}

There is limited research on evaluating model fairness, particularly in the context of pre-trained models \cite{gorrostieta19_interspeech}. Our focus here is on gender, given the insufficient information for other attributes considered in the reference datasets (see Fig. \ref{fig:fairness_gender}). For EMOVOME dataset, models trained using expert labels exhibit a notable bias towards males, as the UA is around 10\% higher for males in valence and arousal prediction and 1.7\% in emotion categories prediction. The use of non-expert labels resulted in an increase in UA for males of 4.6\% valence, but it was 1.9\% higher for females on arousal. Interestingly, the combined label yielded the most similar results for both genders, with 0.3\% for valence and 1.4\% for arousal. In the case of IEMOCAP, again, UA was higher for male speakers in valence (4\%) but lower in arousal (-3.4\%). For categories, the UA for males was +5.9\% compared to females. Finally, RAVDESS presents the highest difference between both genders, with the UA for females being 29.7\% higher for females compared to males. For arousal and valence, the results are also higher for females (1\% and 6.5\%, respectively). 

Overall, SER models obtain better test results for male speakers in EMOVOME, following previous studies in Spanish datasets \cite{duville2021mexican, costantini2022}. In the reference datasets, IEMOCAP aligns with the observed trend, while RAVDESS shows the opposite results. However, it's important to note that both datasets have a limited number of speakers in the test set (two for IEMOCAP and five for RAVDESS), so no significant conclusions can be drawn from these results. 

% ------------------------------------------------------------------------------------------
% Limitations and future work
% ------------------------------------------------------------------------------------------
\subsection{Limitations and future work}

Firstly, while we manually screened for ad hoc recordings, complete control over this factor remains challenging, necessitating reliance on participant integrity. Secondly, improving the annotation process of EMOVOME samples within emotion categories by introducing new raters could alleviate potential individual biases that impact SER models. Similarly, expanding the pool of non-expert labels with more raters may help mitigate bias and achieve a more balanced data distribution regarding arousal. Additionally, other datasets, such as EmoSpanishDB or MOUD, could be explored to increase the number of training samples and assess potential improvements in model performance. Furthermore, in the creation of SER models based on pre-trained models, the current use of average time pooling as an aggregation approach may exhibit suboptimal performance, particularly noticeable in the EMOVOME dataset, where audio duration exhibits significant variability. Future research endeavors will focus on refining time aggregation methods and exploring alternative techniques to address these challenges effectively.

% ##########################################################################################
% Conclusion
% ##########################################################################################
\section{Conclusions}

We developed and publicly released EMOVOME, the first public dataset with realistic emotions from spontaneous conversations in a real-life scenario. We developed speaker-independent SER models using EMOVOME, and compared the results with IEMOCAP and RAVDESS. Superior results were achieved with pre-trained transformer-based models compared to baseline models based on acoustic features. However, EMOVOME results demonstrated lower performance compared to the acted RAVDESS dataset. For the elicited IEMOCAP dataset, the prediction of emotion categories outperformed EMOVOME, but similar results were obtained for predicting valence and arousal. A comprehensive study was also conducted to assess the influence of different properties of EMOVOME on SER performance. Notably, we found variations depending on the labels provided by different annotators, with superior outcomes observed when utilizing combined labels from both expert and non-experts. Interestingly, this combined label also yielded the most equitable results when assessing gender fairness. The results highlight the importance of natural datasets for predicting genuine emotions, emphasizing the need for speech from real-life scenarios to accurately assess the challenges associated with authentic environments.

% ##########################################################################################
% Acknowledgment
% ##########################################################################################
\section*{Acknowledgments}

This work was supported by the European Union’s Horizon 2020 funded project HELIOS (ID 825585), by the Universitat Politècnica de València (PAID-10-20), by the Generalitat Valenciana (ACIF/2021/187), and by the Spanish Ministry of Science and Innovation for the BEWORD project (PID2021-126061OB-C41) and for the DIPSY project (TED2021-131401B-C21).

% ##########################################################################################
% References
% ##########################################################################################

\bibliographystyle{IEEEtran}
\bibliography{main}

\end{document}